\documentclass[twocolumn,amsmath,showkeys,amssymb,superscriptaddress,longbibliography,floatfix, prl]{revtex4-2}
\usepackage{graphicx, subfigure}
\usepackage{algorithm}
\usepackage{algpseudocode}
\usepackage{amssymb, amsmath,amssymb,amsfonts}
\usepackage{amsthm,mathrsfs,amsopn}
\usepackage{dcolumn}
\usepackage{bm}
\usepackage{color}
\usepackage[utf8]{inputenc}
\usepackage{mathtools}
\usepackage[normalem]{ulem}
\usepackage{siunitx} 
\usepackage{amsmath}
\usepackage{color}
\usepackage{bm}

\usepackage[utf8]{inputenc}
\definecolor{green}{rgb}{0.,0.5,0.5}

\definecolor{orange}{rgb}{0.,0.5,0.5}

\definecolor{blue}{rgb}{0.75,0.75,0}
\def\be{\begin{equation}}
\def\ee{\end{equation}}
\def\bc{\begin{center}}
\def\ec{\end{center}}
\def\bea{\begin{eqnarray}}
\def\eea{\end{eqnarray}}
\usepackage{mathtools}

\theoremstyle{plain} 

\definecolor{brickred}{rgb}{0.7, 0.25, 0.33}
\definecolor{applegreen}{rgb}{0.55, 0.71, 0.0}

\begin{document}

\title{Dirac-Equation Signal Processing: Physics Boosts Topological Machine Learning }
\author{Runyue Wang}
\affiliation{School of Mathematical Sciences, Queen Mary University of London, London, E1 4NS, United Kingdom}
\author{Yu Tian}
\affiliation{Nordita, KTH Royal Institute of Technology and Stockholm University, SE-106 91 Stockholm, Sweden}
\affiliation{Center for Systems Biology Dresden, 108 Pfotenhauerstraße 
01307 Dresden, Germany}
\author{Pietro Li\`o}
\affiliation{Department of Computer Science and Technology,
University of Cambridge, Cambridge CB3 0FA, United Kingdom}
\author{Ginestra Bianconi}
\affiliation{School of Mathematical Sciences, Queen Mary University of London, London, E1 4NS, United Kingdom}

\begin{abstract}
Topological signals are variables or features associated with both nodes and edges of a network. Recently, in the context of Topological Machine Learning, great attention has been devoted to signal processing of such topological signals. Most of the previous topological signal processing algorithms treat node and edge signals separately and work under the hypothesis that the true signal is smooth and/or well approximated by a harmonic eigenvector of the Hodge-Laplacian,  which may be violated in practice.  Here we propose  Dirac-equation signal processing, a framework for  efficiently reconstructing true signals on nodes and edges, also if they are not smooth or harmonic, by processing them jointly. The proposed physics-inspired algorithm is based on the spectral properties of the topological Dirac operator. It leverages the mathematical structure of the topological Dirac equation to boost the performance of the signal processing algorithm. We discuss how the relativistic dispersion relation obeyed by the topological Dirac equation can be used to assess the quality of the signal reconstruction. Finally, we demonstrate the improved performance of the algorithm with respect to previous algorithms. Specifically, we show that Dirac-equation signal processing can also be used efficiently if the true signal is a non-trivial linear combination of more than one eigenstate of the Dirac equation, as it generally occurs for real signals. 

\end{abstract}

\maketitle

Physics and Artificial Intelligence are strongly related~\cite{carleo2019machine} as the theory of information is at the core of natural physical systems as well as of learning. Indeed, it is not by chance that the theory of learning has its roots in physically inspired models such as the Hopfield model \cite{hopfield1982neural} strongly related to statistical mechanics of disordered systems \cite{amit1989modeling,mezard1987spin}. In more recent developments of the field, however, not only classical statistical mechanics has become relevant to understanding learning but also high-energy physics~\cite{kimmel1997high}, quantum physics~\cite{glasser2020probabilistic} and network science \cite{bianconi2021higher,battiston2021physics} that are closer to a geometrical and topological interpretation of data.

Topological Signal Processing~\cite{barbarossa2020topological,schaub2021signal,schaub2022signal,sardellitti2021topological} and Topological Machine Learning~\cite{hensel2021survey,hajij2022topological,papillon2023architectures,hajij2020cell}, are currently at the forefront of Artificial Intelligence and combine algebraic topology~\cite{grady2010discrete} and higher-order networks to learning.  At the core of the field, lies Topological Data Analysis \cite{otter2017roadmap,vaccarino2022persistent} that is now one of the principal approaches in computational neuroscience and has been shown to be very successful in extracting important topological information encoded in brain data \cite{petri2013topological,petri2014homological,reimann2017cliques,curto2017can,giusti2015clique}. More recently, growing scientific interest has been addressed in the development of machine learning algorithms for processing and learning topological signals defined on networks (graphs) as well as simplicial complexes. Topological signals are variables associated not only to nodes but also to the edges of a network or higher dimensional simplices of simplicial complexes. 
Topological signals and specifically edges signals are ubiquitous, as they can in general be used to represent fluxes defined on the edges and also vector fields~\cite{barbarossa2020topological} such as currents in ocean \cite{schaub2020random} or speed of wind at a given altitude and different locations on the Earth. Edge signals are also considered to be key for brain research: at the neuronal level, they describe synaptic signals, while at the level of brain regions, there are new proposals to extract and analyze these signals from brain data \cite{faskowitz2022edges,santoro2023higher}.

From the perspective of higher-order networks \cite{bianconi2021higher,battiston2021physics,battiston2020networks,bick2023higher,salnikov2018simplicial}, the study of topological signals greatly enriches the dynamical description of the networks. Indeed topological signals can undergo collective phenomena such as topological synchronization~\cite{millan2020explosive,carletti2024global,wang2024global}, and topological diffusion~\cite{ziegler2022balanced,torres2020simplicial,schaub2020random,muhammad2006control} that display significant differences with their corresponding node based dynamics.

From the perspective of Artificial Intelligence, signal processing of topological signals ~\cite{barbarossa2020topological,schaub2021signal,schaub2022signal,sardellitti2021topological} leads to new connections with Topology that were less significant for graph signal processing \cite{bronstein2017geometric}, and leads to the definition of a new generation of neural network architectures based on topology \cite{roddenberry2021principled}, on topological message passing~\cite{bodnar2021weisfeiler,bodnar2021weisfeiler} and on sheafs~\cite{bodnar2022neural,barbero2022sheaf,duta2024sheaf}. 

Most of the topological signal processing algorithms are based on the properties of the Hodge-Laplacians \cite{horak2013spectra,lim2020hodge} and treat the topological signal on nodes, edges, triangles, and so on separately, one dimension at a time. 
However, the Hodge-Laplacian is not the only topological operator that can be used to treat topological signals. Recently the Topological Dirac operator \cite{bianconi2021higher} has been proposed as the key topological operator that can treat jointly topological signals on nodes and edges exploiting all the information in the data across different dimensions. In this context it has been demonstrated that the Topological Dirac operator can be used to propose Dirac signal processing (DSP) \cite{calmon2023dirac} that outperforms Hodge-Laplacian signal processing when the true signal deviates significantly from a harmonic signal. Following these two works, the Dirac operator has become a new playground not only to test new emergent dynamical properties of networks and simplicial complexes \cite{giambagli2022diffusion,muolo2024three,carletti2024global,calmon2022dirac,calmon2023local} and to perform Topological Data Analysis  tasks~\cite{baccini2022weighted,wee2023persistent,suwayyid2024persistent,suwayyid2024persistentFDS,wang2023persistent,ameneyro2024quantum,lloyd2016quantum} but also to formulate Dirac-based Gaussian processes \cite{alain2023gaussian} and Dirac-based neural networks  \cite{battiloro2024generalized,nauck2024dirac}.

In this work, we propose the Dirac-equation signal processing (DESP) algorithm that can jointly process node and edge signals of a network. This algorithm is based on the mathematical properties of the Topological Dirac equation \cite{bianconi2021topological} that is the generalization to arbitrary lattices of the staggered fermions by Kogut and Susskind and the Dirac-K\"alher fermions defined on lattices \cite{kogut1975hamiltonian,becher1982dirac} and is inspiring further research in theoretical physics~\cite{bianconi2023mass,bianconi2024quantum,delporte2023dirac} and Artificial intelligence \cite{nauck2024dirac}.
The DESP greatly improves the performance of the algorithm with respect to the DSP algorithm proposed in Ref.~\cite{calmon2023dirac}.
Indeed, the use of the eigenstates of the Topological Dirac equation allows us to treat node and edge signals of different scales whose offset can be modulated by learning an additional parameter of the model that plays the role of the mass in the Topological Dirac Equation.
The DESP can be used to reconstruct signals that are not harmonic under very general conditions. In particular, if the true signal is aligned to an eigenstate of the Topological Dirac equation, DESP can be used to efficiently reconstruct the signal, outperforming both the Hodge-Laplacian signal processing and DSP. In this case, the learning of the mass parameter can be done by minimizing the loss of the algorithm or can be achieved by implementing physics insights and looking for the reconstructed signal that more closely obeys the relativistic dispersion relation which characterizes the eigenstates of the Topological Dirac equation.
When processing real topological signals, however often the true signal is not aligned along a single eigenstate of the Dirac equation. In this case, we propose to use the Iterated Dirac-equation Signal processing (IDESP) algorithm that reconstructs the true signal by singling out the eigenstates of the Topological Dirac Equation forming its decomposition, one eigenstate at a time. 

Here the performance of the DESP and the IDESP is validated over both network models and real networks with both synthetic and real data. 
The performance of the physics-inspired DESP and IDESP has greatly improved with respect to the simple DSP, and this research opens the way for further use of the Topological Dirac equation in machine learning.


\section{Background}
\subsection{Topological spinor}
A graph $G=(V,E)$ is formed by a set $V$ of $N_0$ nodes and a set $E$ of $N_1$ edges.  
The dynamical state of a network $G$ is fully determined by the {\em topological spinor} $\psi$  \cite{bianconi2021topological} which comprised both the node and edge topological signals. Mathematically  the topological spinor $\psi$ is given by the direct sum  $\psi=\chi\oplus\phi$ where  indicated by a $0$-cochain $\chi\in C^0$
encoding for the node signals and a $1$-cochain $\phi\in C^1$
encoding for the edge signals. Thus the topological spinor $\psi\in C^0\oplus C^1$  can be represented as the $\mathcal{N}=N_0+N_1$ column vector $\bm\psi\in \mathbb{R}^{\mathcal{N}}$ with $\mathcal{N}=N_0+N_1$ of block structure
\begin{equation}
    \boldsymbol{\psi}  = \left(\begin{array}{c}
    {\bm\chi} \\
    {\bm\phi}
\end{array}\right),
\label{eq:top_spinor}
\end{equation}
with $\boldsymbol{\phi}\in \mathbb{R}^{N_0}$ being  the $N_0$ column vector representing the  node signals  and $\boldsymbol{\chi}\in \mathbb{R}^{N_1}$ being the  $N_1$ column vector representing the edge signals.

\subsection{Hodge Laplacian signal processing }
\subsubsection{The boundary operator and the Hodge Laplacians}
Discrete exterior calculus \cite{grady2010discrete,bianconi2021higher,bick2023higher} allows us to perform discrete differential operator on topological signals that are fundamental to be able to process and filter them.
The exterior derivative $d:C^{0}\to C^1$ maps  node signals to edge signals and encodes the discrete gradient of the node signal. In particular $d\chi$ is a $1$-cochain associating to each edge the difference between the $0$-cochain $\chi$ calculated at its two end nodes, i.e.
\bea
{[d\chi]}_{\ell=[rs]}=\chi_{s}-\chi_r.
\eea
On an unweighted network, the  discrete divergence of the edge signal $d^{\star}:C^{1}\to C^{0}$ maps edge signal into node signal such that  \bea
{[d^{\star}\phi]}_{r}=\sum_{s=1}^{N_0}\phi_{[sr]}-\sum_{s=1}^{N_0}\phi_{[rs]}.\eea
It follows that both of these operators can be encoded by the  boundary matrix ${\bf B}$ is the $N_{0} \times N_1$  matrix defined as
\begin{equation}
    {\bf B}_{r\ell} = \left\{\begin{array}{cl}
        1 & \mbox{ if } \ell=[s,r] ,\\
        -1 & \mbox{ if } \ell=[r,s],\\
        0 & \mbox{ otherwise.}\end{array}
    \right.
    \label{eq:boundary}
\end{equation}
where ${\bf B}^{\top}$ encodes for the discrete gradient and ${\bf B}$ encodes for discrete divergence.
From the boundary operator we can construct two Hodge Laplacians ${\bf L}_{[0]}={\bf B}{\bf B}^{\top}$ also called the graph Laplacian and  ${\bf L}_{[1]}={\bf B}^{\top}{\bf B}$ also called the $1$-Hodge-Laplacian of the network.
The Hodge Laplacians ${\bf L}_{[0]}$ and ${\bf L}_{[1]}$ describe respectively the diffusion from nodes to nodes through edges and the diffusion from edges to edges through nodes.

\subsubsection{Discussion on Hodge Laplacian Signal Processing (LSP) and the challenges that it raises}
In this paragraph, we introduce Hodge-Laplacian signal processing (LSP) which is an umbrella model including both graph signal processing \cite{bronstein2017geometric} and simplicial signal processing ~\cite{barbarossa2020topological,schaub2021signal,schaub2022signal,sardellitti2021topological}.
Suppose we were given a noisy node or edge signal $\bm\theta\in C^n$ with $n\in \{0,1\}$ given by a true signal $\bm\theta$ plus noise, i.e.
\begin{equation}
    \boldsymbol{\tilde{\theta}} = \boldsymbol{\theta} + \boldsymbol{\epsilon},
\end{equation}
where $\boldsymbol{\epsilon}$ is the noise usually assumed to given by i.i.d. variables associated to each node (for $n=0$) or each edge (for $n=1$). 
For $n=0$, the Hodge-Laplacian signal processing (LSP) assumes that the true node signal is smooth, and thus is formed predominantly by low eigenmodes of the graph Laplacian \cite{bronstein2017geometric}. Similarly, for $n=1$, LSP~\cite{barbarossa2020topological,schaub2021signal,schaub2022signal,sardellitti2021topological} assumes that the true edge signal is almost harmonic, and thus able to capture fluxes going around the holes of the network.
Under these assumptions, the Hodge-Laplacian signal processing allows to generate a reconstructed signal $\hat{\bm\theta}$ that minimizes the loss function $\mathcal{L}_{HL}$
\begin{equation}
    \mathcal{L}_{HL}=\|\hat{\bm\theta}-\tilde{\bm\theta}\|_2^2-\tau \hat{\bm\theta}^{\top}{\bf L}_{[n]} \hat{\bm\theta}.
\end{equation}
Hodge Laplacian signal processing is attracting significant attention for its ability to efficiently reconstruct almost harmonic true signal on networks. Moreover, its extension to higher-order topological signals  allows the treatment of almost harmonic topological signals of higher dimension, i.e. defined also on higher dimensional simplices and cell complexes ~\cite{barbarossa2020topological,schaub2021signal,schaub2022signal,sardellitti2021topological}.

However, the Hodge Laplacian signal processing also has important limitations. On one side, it cannot be used to reconstruct true signals that deviate strongly from harmonic signals. This is relevant because, while for diffusing signals smoothness is a natural assumption, in general, if we consider topological signals that correspond to real features associated to the nodes and edges of a network, we cannot always assume that the signal is smooth or close to harmonic.
The other limitation of this approach is that Hodge Laplacian signal processing treats separately node and edge signals while treating node and edge signals jointly might in principle contribute to reducing the error in the reconstructed signal.
In order to address these two important limitations we will need to use a regularization kernel defined in terms of the topological Dirac operator, defining first the Dirac signal processing and then further improving on this latter algorithm with the Dirac-equation signal processing inspired by theoretical physics.

\subsection{Dirac signal processing (DSP)}
Dirac signal processing (DSP) has been recently introduced in Ref.~\cite{calmon2023dirac} in order to jointly process noisy node and edge signals defined on a network. The algorithm can also be generalized to treat more general signal processing problems defined on simplicial complexes. The key idea of DSP  is to reconstruct the true signal by minimizing a loss function that depends on the Dirac operator \cite{bianconi2021topological} rather than just on the Hodge Laplacian. This key idea is shown to be central in order to efficiently filter the noise from true signals that are not harmonic. In order to introduce DSP let us first discuss the major properties of the Dirac operator.

\subsubsection{Dirac operator}
The Dirac operator ${\bf D}:C^0\oplus C^1\to  C^0\oplus C^1$ \cite{bianconi2021topological} is a differential operator that maps topological spinors into topological spinors and allows topological signals of nodes and edges to cross-talk.  On a network $G$, the Dirac operator $D$ is defined as $D=d+d^{\star}$ and thus the matrix representation ${\bf D}$ of the Dirac operator is a $\mathcal{N}\times \mathcal{N}$ matrix with the following block structure:
\begin{equation}
    {\bf D} = \left(\begin{array}{cc}
        {\bf 0} & {\bf B} \\
        {\bf B}^\top & {\bf 0} 
    \end{array}\right),
\end{equation}
where the boundary operator is defined in Eq.~\ref{eq:boundary}.
Interestingly, the Dirac operator allows topological signals of different dimensions to cross-talk as it is apparent from evaluating  the action of the Dirac operator on the general topological spinor $\bm\psi$ given by Eq.~\eqref{eq:top_spinor}. Indeed we have 
\bea
 {\bf D}\bm\psi=\left(\begin{array}{c}
         {\bf B}\bm\phi \\
        {\bf B}^\top\bm\chi
    \end{array}\right),
    \label{eq:Dirac}
\eea
thus the Dirac operator allows to project node signals into edge signals and edge signals into node signals.
The constitutive property of the Dirac operator is that its square is given by the Gauss-Bonnet Laplacian, i.e.
\begin{equation}
    {\bf D}^2 = \left(\begin{array}{cc}
          {\bf L}_{[0]} &{\bf 0} \\
         {\bf 0}&{\bf L}_{[1]} 
    \end{array}\right).
    \label{eq:Gauss_Bonnet_Laplacian}
\end{equation}
Thus the Dirac operator can be interpreted as the {\em square root of the Laplacian}. Therefore the Dirac operator has a zero eigenvalue with degeneracy equal to the sum of the Betti numbers $\beta_0+\beta_1$, and there is always a basis in which the harmonic eigenvectors are localized only on nodes or on edges. Moreover, since ${\bf L}_{[0]}$ and ${\bf L}_{[1]}$ are isospectral, the non-zero eigenvalues $\lambda$  of the Dirac operator are given by 
\bea
\lambda=\pm\sqrt{\mu},
\eea
where $\mu$ is the generic non-zero eigenvalue of the graph Laplacian ${\bf L}_{[0]}.$ The  eigenvectors associated to eigenvalue $\sqrt{\mu}$ and eigenvalue $-\sqrt{\mu}$ are related by chirality (see for instance discussion in Refs.~\cite{bianconi2021topological,calmon2023dirac}), thus if $(\bm\chi,\bm\phi)$ is associated to the positive eigenvalue,  $(\bm\chi,-\bm\phi)$ is associated to the opposite eigenvalue.
Thus,  the structure of the eigenvectors of the Dirac operator associated to eigenvalues of increasing values (from negative, to zero, to positive) is given by the eigenvector matrix, 
\bea
    &\bm \Phi = 
    \left(\begin{array}{cccc}
    {\mathbf U}  & {\mathbf{0}} & {\bf U}_{\textrm{harm}} & {\mathbf U}\\
    -{\mathbf V} &  {\bf V}_{\textrm{harm}} & {\mathbf{0}} & {\mathbf V}
    \end{array}\right),      
\eea
where ${\bf U}$ and ${\bf V}$ are the matrices of left and right singular vectors of the boundary operator associated to its non-zero singular values, while ${\bf U}_{\textrm{harm}}$ and ${\bf V}_{\textrm{harm}}$ are the matrices of left and right singular vectors of the boundary operator associated to its zero singular values.
In particular, we note that the non-harmonic eigenmodes of the Dirac operator associated to the eigenvalue $\lambda$ enforce  that the node signal $\chi$ is related to the edge signal $\phi$ by $\lambda\bm\phi={\bf B}^{\top}\bm\chi$ and vice versa $\lambda\bm\chi={\bf B}\bm\phi.$ Thus node and edge topological signals of single eigenmodes of the Dirac operator need to have a compatible  normalization, and are not allowed to have arbitrarily different scales.

\subsubsection{Discussion on DSP and the challenges that it raises}
The key idea of DSP introduced in Ref.~\cite{calmon2023dirac} is to process jointly node and edge signals in order to be able to exploit all the relevant information present in the topological spinor.
We assume that the true data is encoded by the topological spinor $\boldsymbol{\psi}$, but that we have only access to the noisy signal $\boldsymbol{\tilde{\psi}}$ given by
\begin{equation}
    \boldsymbol{\tilde{\psi}} = \boldsymbol{\psi} + \boldsymbol{\epsilon},
\end{equation}
where ${\bf \epsilon}$  indicates the noise. 
As we have seen in the previous chapters, the underlying assumption of LSP is that the true signal is harmonic, or close to harmonic. On the contrary, in DSP the underlying assumption is that the signal has a major contribution aligned with the eigenvector associated to  the eigenvalue $\lambda=E$ of the Dirac operator, where the exact value of $E$ can be actually learned by the algorithm.
Given the noisy signal $ \boldsymbol{\tilde{\psi}}$,  DSP  reconstructs the signal $\bm{\hat{\psi}}$ by minimizing the loss function,
\begin{equation}
    \mathcal{L}=\|\boldsymbol{\hat{\psi}}-\boldsymbol{\tilde{\psi}} \|_2^2 + \tau\boldsymbol{\hat{\psi}}^\top(\boldsymbol{D} - E{\bf I})^2 \boldsymbol{\hat{\psi}},
\end{equation}
where $E\in \mathbb{R}$ and ${\bf I}$ indicates the identity matrix. The regularization term $\bm{\mathcal{R}}=\boldsymbol{\hat{\psi}}^\top({\bf D} - E{\bf I})^2\boldsymbol{\hat{\psi}}$ filters more the components of the measured signal associated to an eigenvalue $\lambda$ of the Dirac operator ${\bf D}$ that depart more significantly from $E$, i.e., for which $(\lambda-E)^2$ is large. Note however that the parameter $E$ is not an external input of DSP algorithm and can be learned by the algorithm under very general conditions~\cite{calmon2023dirac}.

It is also instructive to consider the limit in which $E=0$, i.e., the true signal is indeed almost harmonic. In this case, the loss $\mathcal{L}$ reduces to 
\begin{equation}
    \mathcal{L}=\| \boldsymbol{\hat{\psi}}-\boldsymbol{\tilde{\psi}}\|_2^2 + \tau \boldsymbol{\hat{\psi}}^\top  {\bf D}^2 \boldsymbol{\hat{\psi}}.
    \label{eq:DSP}
\end{equation}
and since ${\bf D}^2$ is the Gauss-Bonnet Laplacian defined in Eq.~\ref{eq:Gauss_Bonnet_Laplacian}, it follows that DSP in this limit reduces to the LSP treating node and edge signals independently. 

Dirac signal processing has been shown~\cite{calmon2023dirac} to have an excellent performance when the true signal is an eigenstate of the Dirac operator, while when it is applied to true data the accuracy of the signal reconstruction decreases. 
Here we identify two reasons for this decrease in the performance on real data. One reason is that the non-harmonic eigenmodes of the Dirac operator imply a strict relation between the norm of the node signal and the norm of the edge signal, while on real data node and edge signals might have a different scale. The second reason is that the true signal might be given by the combination of more than two eigenmodes of the Dirac operator. 
In order to address these two limitations, in this work we propose the Dirac-equation signal processing and the Iterated Dirac-equation signal processing that greatly improves the performance of the Dirac signal processing on real data.

\section{Dirac-equation signal processing (DESP)}
Here we introduce the Dirac-equation signal processing (DESP), a signal processing algorithm that can jointly process node and edge signals that reduces to LSP and to DSP in limiting cases and in the most general case can overcome the limitations of the previously discussed signal processing algorithms. The formulation of the DESP is inspired by theoretical physics and builds on the mathematical structure of the eigenstates of the Topological Dirac equation~\cite{bianconi2021topological}. 
Thus, before  discussing the DESP algorithm and its performance on synthetic and real data, let us first outline the main properties of the Topological Dirac equation.
\subsection{Topological Dirac equation}
The  Topological Dirac equation \cite{bianconi2021topological} is a differential equation for a quantum wave function defined on an arbitrary network. This equation is the natural extension to an arbitrary network of the staggered fermions by Kogut and Susskind \cite{kogut1975hamiltonian} and the Dirac-K\"ahler fermions \cite{becher1982dirac} defined on a lattice. 
The Dirac equation is a wave equation for the topological spinor,  defined as
\begin{equation}
    \textrm{i} \partial_t \boldsymbol{\psi} = \boldsymbol{\mathcal{H} \psi},
\end{equation}
where  the Hamiltonian $\boldsymbol{\mathcal{H}}$  is linear on the Dirac operator ${\bf D}$ and depends on the mass $m\geq 0$ as
\begin{equation}
    \boldsymbol{\mathcal{H}} = {\bf D} + m \boldsymbol{\gamma},\end{equation}
    with the matrix $\bm\gamma$ being given by
    \begin{equation}
    {\bm\gamma} = \left(\begin{array}{cc}
        {\bf I}_{N_0} & {\bf 0} \\
        {\bf 0} & -{\bf I}_{N_1} 
    \end{array}\right).
    \label{eq:gamma}
\end{equation}
The eigenstates $\boldsymbol{\psi}$ of the Topological Dirac equation associated to energy $E$ satisfy the eigenvalue problem 
\begin{equation}
    E\boldsymbol{\psi} = ({\bf D} + m \boldsymbol{\gamma}) \boldsymbol{\psi}.
\end{equation}
Using the definition of the Dirac operator Eq.~\ref{eq:Dirac} and the definition of the gamma matrix ${\bm\gamma}$ Eq.~\ref{eq:gamma} this eigenvalue system can be written as 
\bea
    E\boldsymbol{\phi}&=&{\bf B}\bm\chi+m{\bm\phi},\nonumber\\
    E\boldsymbol{\chi}&=&{\bf B}^\top{\bm\phi}-m{\bm\chi}.
    \eea
Thus, after a few algebraic steps we get
\bea
    (E-m)(E+m) \boldsymbol{\phi}={\bf BB}^\top{\bm \phi}={\bf L}_{[0]}\boldsymbol{\phi},\nonumber\\
    (E-m)(E+m)\boldsymbol{\chi}={\bf B}^\top{\bf B}{\bm\chi}= {\bf L}_{[1]}\boldsymbol{\chi}.
\eea
This implies that the node signal $\bm\chi$ is an eigenvector of the graph Laplacian ${\bf L}_{[0]}$ with eigenvalue $\mu=\lambda^2$ and that the edge signal $\bm\phi$ is an eigenvector of the $1$-Hodge Laplacian ${\bf L}_{[1]}$ with the same eigenvalue, where the energy $E$ is related to $\lambda$ through the {\em relativistic dispersion relation}
\begin{equation}
    E^2=m^2+\lambda^2.
\end{equation}
In particular, it can be shown that both positive and negative energy states are realized with 
\bea
E=\pm \sqrt{m^2+\lambda^2}.
\label{eq:relativistic_dispersion_relation}
\eea
Thus the role of the mass is to introduce a gap in the energy spectrum, as the energy values need to have an absolute value greater or equal to the mass, i.e.~$|E|\geq m$.
\begin{figure*}[hbt!]
    \centering
        \includegraphics[width=\textwidth]{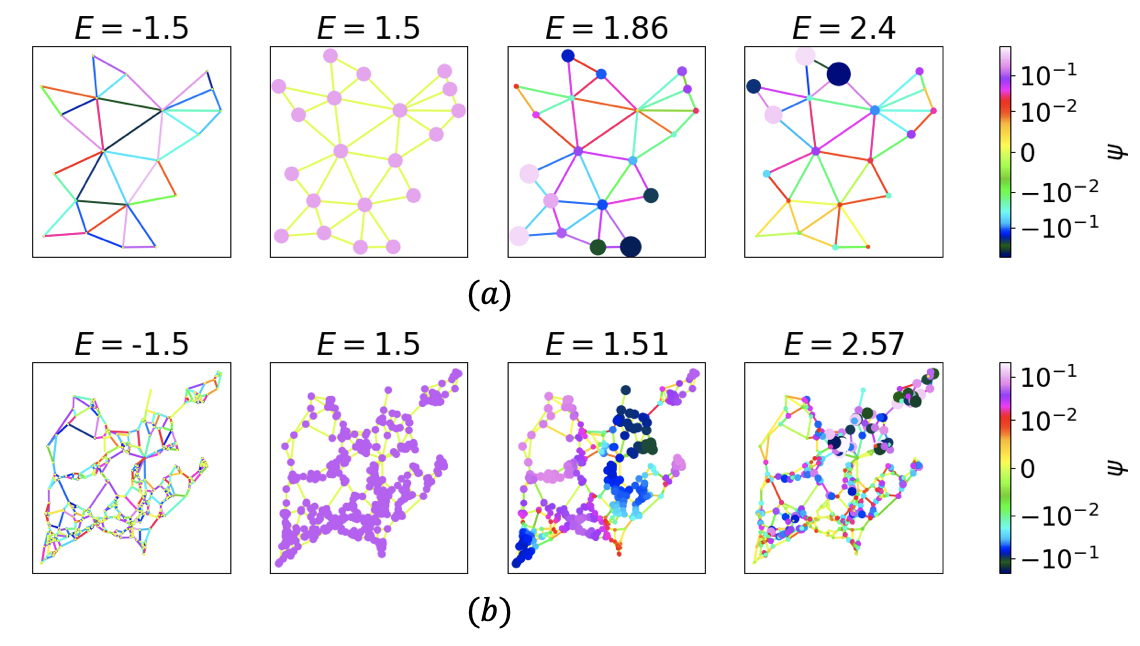}
    \caption{
    The visualization of the eigenstates of the topological Dirac equation associated with the value of the mass $m=1.5$ on the Network Geometry with Flavor model (NGF) (panel (a)) and on a real fungi network (panel (b)). The eigenstates $E=m=-1.5$ and $E=m=1.5$ are the harmonic eigenstates which are non-zero only on edges ($E=-m=-1.5$) or only on nodes ($E=m=1.5$). The eigenstates with energy $E>m=1.5$ are non-harmonic,  they involve non trivially both node and edge signals, and display characteristic localized patterns. These latter signals are typical examples of signal that can be reconstructed with the DESP. The NGF network in panel (a) is a sample of a  two dimensional NGF model with parameters $\beta=0$ and flavor $s=-1$ $N_0=20$ nodes and $N_1=37$ edges. This model is defined in Refs.~\cite{bianconi2016network,bianconi2017emergent} and the code for generate network in this model is available at the repository \cite{gin_repository}. The  fungi network in panel (b) is the  \texttt{Pp\_M\_Tokyo\_U\_N\_26h\_1.mat}, of $N_0 = 411$ nodes, and $N_1 = 645$ edges  from Ref.~\cite{lee2017mesoscale} available at the repository \cite{benson_repository}. }
    \label{fig0}
\end{figure*}
The mass changes also significantly the properties of the eigenstates associated to non-harmonic eigenvectors $\lambda>0$.
In order to see this, let us discuss the  structure of the eigenvectors, encoded in the matrix of eigenvectors $\bm\Phi$
\bea
    &\bm \Phi = 
    \left(\begin{array}{cccc}{\bm \Psi}^{-} &  {\bm \Psi}_{\textrm{harm}}^{-}&{\bm \Psi}_{\textrm{harm}}^{+}&{\bm \Psi}^{-}  \\
    \end{array}\right).   
\eea
Here ${\bm \Psi}^{\pm}$ are the matrices associated to  the eigenvectors with $\lambda\neq 0$ and $E>m$ or $E<-m$, respectively,  which are given by 
\bea
    \boldsymbol{\psi}_{\lambda}^+ = \mathcal{C}^+\left(\begin{array}{c}
        {\bf u}_\lambda \\
        \frac{ \lambda}{|E|+m} {\bf v}_\lambda
    \end{array}\right),\quad
    \boldsymbol{\psi}_{\lambda}^- = \mathcal{C}^- \left(\begin{array}{c}
    \frac{ \lambda}{|E|+m} {\bf u}_\lambda \\
       -{\bf v}_\lambda
    \end{array}\right),
    \eea
    where ${\bf u}_\lambda$ and ${\bf v}_{\lambda}$ are the  left and right singular vectors of the boundary operator ${\bf B}$ associated to the singular value $\lambda$ and $\mathcal{C}^{\pm}$ are normalization constants.
    We note that the mass allows now to tune the relative normalization of the node and the edge signal which can now have very different scales. Only for  $m=0$ these eigenvectors reduce to the eigenvector of the Dirac operator. 
 The eigenvectors that are associated to $\lambda=0$ and energy $E=\pm m$ are instead the harmonic eigenvectors. These eigenvectors  are independent of the value of the mass and are given by  
\bea
    \boldsymbol{\psi}_{\textrm{harm}}^{+} =  \left(\begin{array}{c}
        {\bf u}_0 \\
        {\bf 0}
    \end{array}\right),\quad
    \boldsymbol{\psi}_{\textrm{harm}}^{-} =  \left(\begin{array}{c}
        {\bf 0} \\
        {\bf v}_0 \\
    \end{array}\right).
\eea
Note that the degeneracy of the eigenvalue $E=m$ is given by the $0$-Betti number $\beta_0$, while the degeneracy of the eigenvalue $E=-m$ is given by the $1$-Betti number $\beta_1$.
 In Figure $\ref{fig0}$, we represent the eigenstates of the topological Dirac equation on two different networks: the network skeleton of the Network Geometry with Flavor (NGF) model \cite{bianconi2016network,bianconi2017emergent} and a real fungi network from Ref.~\cite{lee2017mesoscale}.
 From this figure, it is apparent that the harmonic eigenstates with energy $E=\pm m$ are significantly different from the non-harmonic eigenstates $|E|>m$. Indeed the harmonic eigenstates are non-trivially defined only on the nodes ($E=m$) or only on the edges ($E=-m$) with the harmonic mode at $E=m$ being constant on the nodes and the generic harmonic mode at $E=-m$ being a linear combination of modes localized on the cycles of the network. However the non-harmonic eigenstates of the topological Dirac equation at $|E|>m$ involve non-trivial pattern localization and non-trivial distribution of the signal on both nodes and edges.
 It is clear that in general, an arbitrary topological network signal might not be harmonic, thus formulating a signal processing algorithm to infer these signals is an important research question.

\subsection{DESP: Problem set up and algorithm}
\label{sec:DESP}
Considering a noisy topological signal $\bm{\tilde{\psi}}$ defined on both nodes and edges and given by the true signal ${\bm\psi}$ of the unitary norm, i.e.~$\|\bm\psi\|_2=1$,  plus the noise ${\bm\epsilon}$, i.e.
\bea
\bm{\tilde{\psi}}=\bm{{\psi}}+{\bm\epsilon},
\eea
where $\bm\epsilon$  indicates the noise with noise level $\alpha$ (see Methods for details).
The DESP aims at reconstructing the true signal making minimal assumptions.
The assumption of the DESP is that the true signal is a general eigenvector of the Topological Dirac equation with energy $E$ and mass $m$ to be determined by the algorithm where here and in the following. 
For $E=m=0$, this assumption coincides with the underlying assumption of LSP, i.e.~that the signal is harmonic or close to harmonic, and indeed the DESP algorithm reduces to LSP in this case. For $m=0$, this assumption coincides with the underlying assumption of DSP that the topological signal can be a general eigenmode of the Dirac operator, and indeed in this limit we recover DSP.
However in the general case where $E\neq0, m\neq 0$, DESP cannot be reduced to any of the previous algorithms and displays a much better performance for general signals than the previous two algorithms as it allows node and edge signal to have a different scale.
Interestingly it is to be noted that the DESP admits a variation, the Iterative Dirac-equation signal processing (IDESP) that would allow us in the next section to go even beyond the assumption that the true signal is aligned to a single eigenstate of the Topological Dirac equation and to reconstruct efficiently true signals that are linear combinations of different eigenstates of the Topological Dirac equation that occur in real data.
In DESP the reconstructed signal $\boldsymbol{\hat{\psi}}$ is obtained by minimizing the following loss function:
\begin{equation}
    \mathcal{L}(\boldsymbol{\hat{\psi}})=   \| \boldsymbol{\hat{\psi}} - \boldsymbol{\tilde{\psi}} \|_2^2 + \tau \boldsymbol{\hat{\psi}}^\top  (\boldsymbol{D} +m\boldsymbol{\gamma} - E{\bf I} )^2 \boldsymbol{\hat{\psi}},
\end{equation}
where here and in the following we use the notation ${\bf I}={\bf I}_{\mathcal{N}}$.
Note that here the regularization term leaves unchanged the component of the noisy signal aligned to the eigenstate of the Topological Dirac equation with energy $E$ and mass $m$ while filtering out components associated with an energy $E^{\prime}$ that deviates from $E$ with a filter proportional to $(E^{\prime}-E)^2$.
For $m=0$, we get the loss function of DSP given by Eq.~\ref{eq:DSP}, and when also $E=0$, the algorithm reduces to the two decoupled LSP algorithms for node and edge signals.
The significant benefit to considering DESP with respect to DSP is the fact that by introducing the mass $m$, DESP allows us to treat efficiently topological spinor whose node and edge signals have different scales as it occurs in general in data.
The loss function can be minimized with respect to the reconstructed signal $\boldsymbol{\hat{\psi}}$
obtaining 
\bea
\boldsymbol{\hat{\psi}}=\left[{\bf I}+\tau (\boldsymbol{D} +m\boldsymbol{\gamma} - E{\bf I} )^2\right]^{-1}\bm{\tilde{\psi}}.
\eea
Moreover, the loss $\mathcal{L}$ can also be minimized with respect to  $E$ and $m$ getting 
 \bea
    m =  \frac{\boldsymbol{\hat{\psi}}^\top ( E{\bf I}-{\bf D}) \boldsymbol{\hat{\psi}}}{\boldsymbol{\hat{\psi}}^\top \boldsymbol{\gamma}\boldsymbol{\hat{\psi}}},\quad
    E = \frac{\boldsymbol{\hat{\psi}}^\top ({\bf D} + m\boldsymbol{\gamma}) \boldsymbol{\hat{\psi}}}{\boldsymbol{\hat{\psi}}^\top\boldsymbol{\hat{\psi}}}.
\eea
Note that for the purpose of the DESP we will allow the mass $m$ to take also negative real values as this is allowed in this topological setting (it is equivalent to changing the sign in front of the $\bm\gamma$ matrix).
\begin{figure}[!htb!]
    \centering
    \begin{tabular}{c}
        \includegraphics[width=0.95\columnwidth]{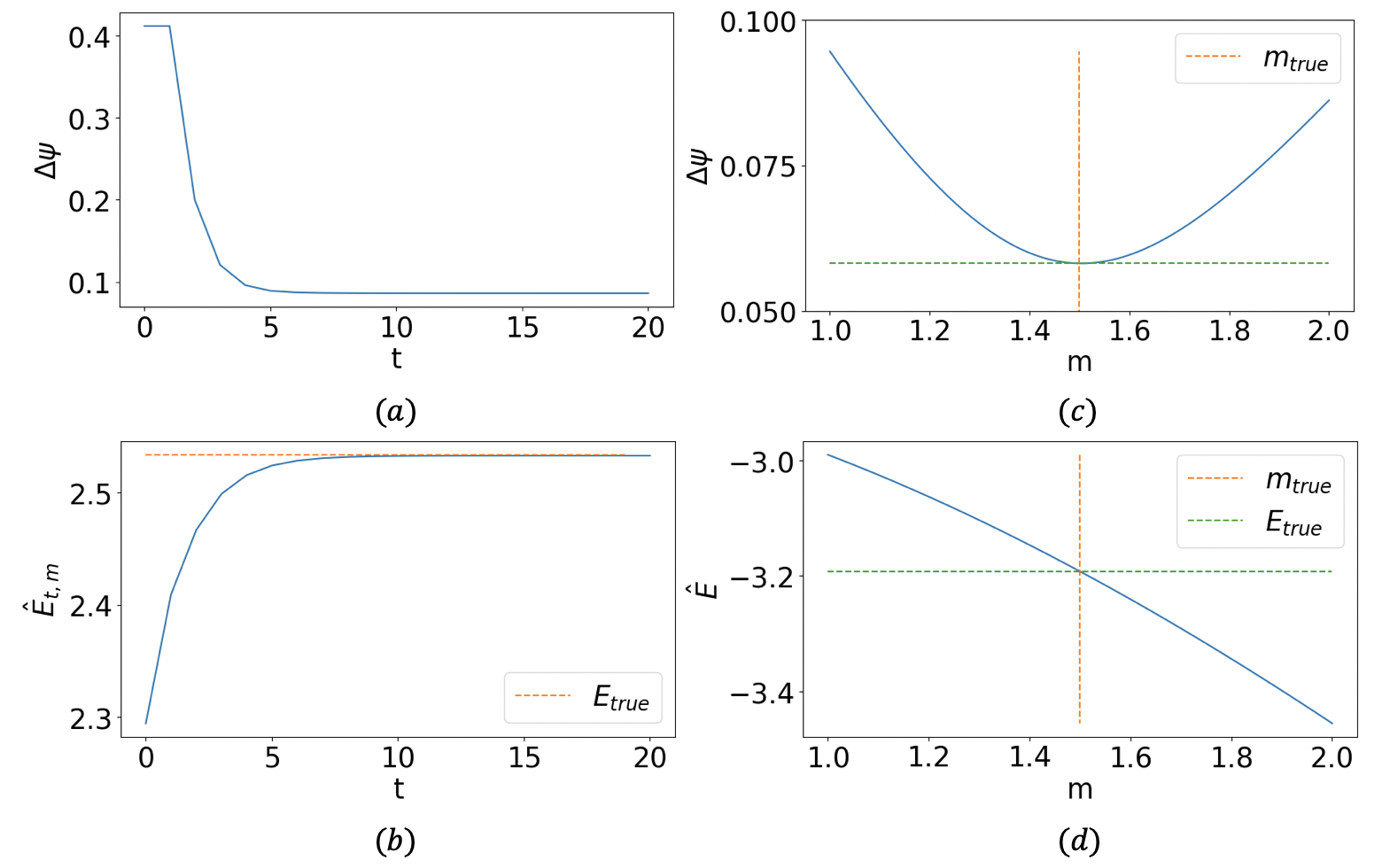}
    \end{tabular}
    \caption{
    We consider a true signal given by an eigenstate of the Topological Dirac equation. For any fixed value of $m$ the iterative nature of the DESP algorithm allows to decrease the true error $\Delta \bm\psi$ with time $t$ (panel (a)) and to best approximate the energy of the signal with time $t$ so that if $m$ is the true mass $m=m_{\textrm{true}}$ then the estimated energy $\hat{E}_{t,m}$ converges to the true energy $E_{\textrm{true}}$ (panel (b)). As the algorithm sweeps over different values possible value of the mass $m$, the true value of the mass and the true value of the energy are reliably estimated under very general conditions on the noise level (panels (c) and (d)).
    Here the DESP convergence to the true mass and energy parameters is demonstrated on the NGF network shown in Figure 1. The true value of the mass is $m_{\textrm{true}}=1.5$ and the true value of the energy is $E_{\textrm{true}}=-3.19$.
    The noise is generated using a value of the $\alpha$ parameter given by  $\alpha = 0.3$, while the loss $\mathcal{L}$ used to detect both the energy and mass has filtration parameter $\tau = 10.$ } 
    \label{fig:dsp-1d-errE}
\end{figure}
Theoretically, it is possible to optimize $ \boldsymbol{\hat{\psi}}, m , E$ simultaneously. However, we would also like to guarantee computational efficiency, with a cost of negligible difference in accuracy. 
The  DESP Algorithm (see pseudo-code in Algorithm \ref{alg:DESP}) considers a sweep over different values of $m$, where in practice the values of the mass $m$ will span an interval bounded by the extrema of eigenvalues of the Dirac operator ${\bf D}$. For each value of $m$, the DESP algorithm optimizes the reconstructed signal $\bm{\hat{\psi}}_{m}$ and learns the best value of the reconstructed energy $\hat{E}_m$. This is done by iteratively interpolating the value of the estimated energy with the estimated value of the energy that minimizes the loss function. This iterative optimization is performed using the Armijo rule \cite{Armijo66Math} that ensures that the interpolation parameter is chosen in such a way to guarantee the decrease of the loss function at each step of the iteration. 
Having performed the sweep over the relevant values of the mass, the best choice of the mass can be selected according to different criteria. The default possibility is to minimize the loss $\mathcal{L}$ calculated over the reconstructed signal $\bm{\hat{\psi}}_{m}$ and energy $\hat{E}_m$, associated to the mass $m$, i.e.~minimizing $\mathcal{L}_m$ given by \bea
\mathcal{L}_m= \| \boldsymbol{\hat{\bm\psi}}_m - \boldsymbol{\tilde{\bm\psi}} \|_2^2 + \tau \boldsymbol{\hat{\bm\psi}}_m^\top  (\boldsymbol{D} + m\boldsymbol{\gamma} - E{\bf I} )^2 \boldsymbol{\hat{\bm\psi}}_m.
\label{eq:loss_m}
\eea
Thus the reconstructed signal $\bm{\hat{\psi}}$ is the reconstructed signal $\bm{\hat{\psi}}_m$ corresponding to the optimized value of the mass $m$. Note that alternatively, we can optimize the value of the mass using the relativistic dispersion relation as we will discuss in the next paragraph.
\begin{algorithm}[H]
	\caption{Dirac-Equation Signal Processing (DESP) }
	\label{alg:DESP}
	\begin{algorithmic}[1] 
 \Statex Inputs:   noisy signal $\boldsymbol{\tilde{\psi}}$,  Dirac operator ${\bf D}$,  gamma matrix $\boldsymbol{\gamma}$, the regularization parameter $\tau$,  initial learning rate $\sigma,$  precision $\delta E$ for inferred energy values $\hat{E}$, minimum number of iterations $T$, interval of possible mass values $m\in [0,\bar{m}]$ with $\bar{m}>0$,  precision $\delta m $ for any inferred mass value $m$, method used to find optimal value of the mass, choose from: loss function optimization $\mathcal{L}$, (default), relativistic dispersion relation method. 
 \Statex Output: The output of the DESP Algorithm starting from the generic noisy signal $\tilde{\bm\psi}$ is indicated as $\hat{\bm\psi}=\textrm{DESP}(\tilde{\bm\psi})$.
\State $m  \leftarrow 0$
\While{$m\leq \bar{m}$}
            \State $t\leftarrow 0$
            \State $\boldsymbol{\hat{\psi}}_{t,m} \leftarrow \boldsymbol{\tilde{\psi}}$
            \State $\hat{E}_{t,m} \leftarrow \frac{\boldsymbol{\tilde{\psi}}^\top ({\bf D} + m\boldsymbol{\gamma}) \boldsymbol{\tilde{\psi}}}{\boldsymbol{\tilde{\psi}}^\top\boldsymbol{\tilde{\psi}}}$ 
            \While{$| \hat{E}_{t,m} - \hat{E}_{t - 1,m}|> \delta E$ or $t < T$}
    		\State 
                $\boldsymbol{\hat{\psi}}_{t+1,m} \leftarrow \left[{\bf I} +  \tau  (\boldsymbol{D} + m\boldsymbol{\gamma} - \hat{E}_{t,m}{\bf I})^2\right]^{-1}  \boldsymbol{\tilde{\psi}}$
    \State 
        $\hat{E}_{t+1,m} \leftarrow (1-\sigma) \hat{E}_{t,m} + \sigma \frac{\boldsymbol{\hat{\psi}}_{t+1,m}^\top ({\bf D} + m\boldsymbol{\gamma}) \boldsymbol{\hat{\psi}}_{t+1,m}}{\boldsymbol{\hat{\psi}}_{t+1,m}^\top\boldsymbol{\hat{\psi}}_{t+1,m}},$
    where $\sigma$ follows  Armijo's rule.      
  \State $t \leftarrow t + 1.$
            \EndWhile
    \State $\boldsymbol{\hat{\psi}}_{m} \leftarrow \boldsymbol{\hat{\psi}}_{t,m}$
    \State $\hat{E}_{m} \leftarrow {\hat{E}}_{t,m}$
\State ${\mathcal{L}}_{m} \leftarrow  \| \boldsymbol{\hat{\psi}}_m - \boldsymbol{\tilde{\psi}} \|_2^2 + \tau \boldsymbol{\hat{\psi}}_m^\top  (\boldsymbol{D} + m\boldsymbol{\gamma} - \hat{E}_m{\bf I} )^2 \boldsymbol{\hat{\psi}}_m$
\State $\mathcal{S}_m\leftarrow \left|\frac{\left(\hat{\psi}_m^{\top}({\bf D}+m\bm\gamma)\hat\psi_m\right)^2}{\|\hat{\psi}_m\|^4}-\frac{\hat{\psi}_m^{\top}{\bf D}^2\hat\psi_m}{\|\hat{\psi}_m\|^2}-m^2\right|$
 \State $m\leftarrow m+\delta m$
            \EndWhile
            \State $\hat{m}\leftarrow \mbox{argmin}_{m} \mathcal{L}_m$
            \State ${m}_{\mathcal{S}}\leftarrow\mbox{argmin}_{m} \mathcal{S}_m$
            \State If inferring  $m$ by minimizing  $\mathcal{L}_m$ (default): ${\bm{\hat{\psi}}}={\bm\psi_{\hat{m}}}$ 
             \State If inferring  $m$ by minimizing $\mathcal{S}_m$: 
            $\boldsymbol{\hat{\psi}}={\bm\psi_{m_\mathcal{S}}}$
	\end{algorithmic}
\end{algorithm}
If the true signal $\bm\psi$ is known, the performance of the  DESP algorithm for every value of the mass $m$ can be directly evaluated by calculating the error $\Delta_m \bm\psi$  given by 
\bea
\Delta_m \bm\psi=\|\hat{\bm\psi}_m-\bm\psi||_2.
\eea
where $\hat{\bm\psi}_m$ is the reconstructed signal assuming the mass of $m$.
Finally the error made by the DESP is given by $\Delta \bm\psi$ given by 
\bea
\Delta \bm\psi=\|\hat{\bm\psi}-\bm\psi||_2.
\eea
In Figure $\ref{fig:dsp-1d-errE}$, we show the performance of the  DESP algorithm when the true signal is aligned to a single eigenstate of the Topological Dirac equation under very general conditions on the noise level.  For each value of $m$ considered by the algorithm, the iteration procedure lowers the error $\Delta \bm\psi$ (panel (a)) and finds the energy that best approximates the true energy (panel (b)). In particular if $m$ is given by the true value $m_{\textrm{true}}$, the energy $\hat{E}_{t,m}$ converges to the true energy value ${E}$ as the number of iterations increases (panel (b)). 
Moreover, if we do not know the value of the true mass, by performing the sweep over $m$, the algorithm can efficiently recover the true value of the energy $E$ and the mass $m$  (panel (c) and (d)).
\begin{figure*}
    \centering
  \includegraphics[width=0.99\textwidth]{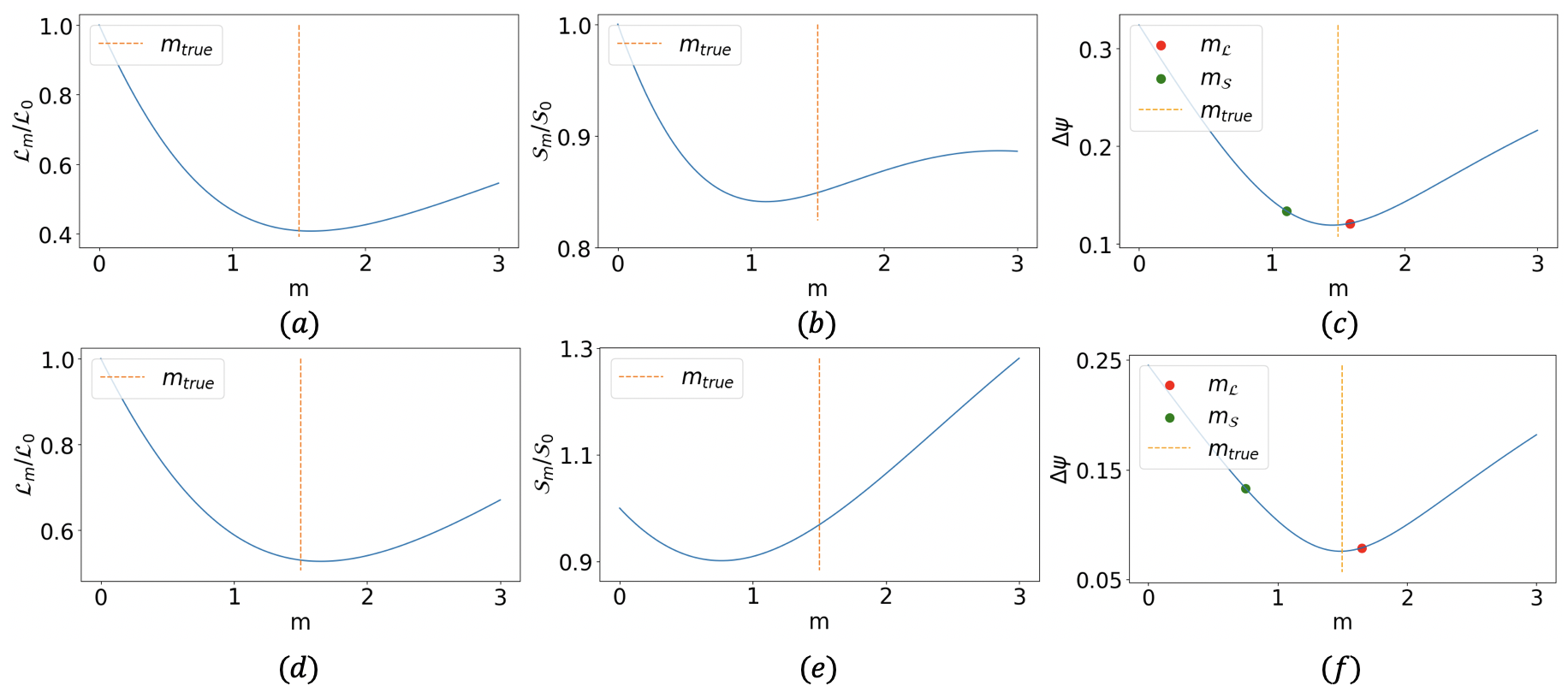}
    \caption{The loss $\mathcal{L}_m$, and the RDRE $\mathcal{S}_m$, are plotted versus $m$ when the true signal is aligned to a single eigenstate of the Topological Dirac equation of the NGF (panels (a)-(b)) and the fungi network (panels (d)-(e)) considered in Figure 1. The minimization of the loss $\mathcal{L}_m$ and of the RDRE $\mathcal{S}_m$ lead to different estimated values of the mass (panels (c), (f)). However, the error $\Delta \bm\psi$ corresponding to these two methods to infer the true mass remains small under very general conditions on the noise level (panels (c), (f)).   The Topological Dirac equation eigenstates have true parameters $E_\textrm{true}=-3.27$ and $m_{\textrm{true}}=1.5$ for the NGF  (panels (a)-(c)) and $E_\textrm{true}=2.57$ and $m_{\textrm{true}}=1.5$ for the fungi network  (panels (d)-(f)). The noise level is  $\alpha = 0.3$ and $\tau = 10$ for both cases.}
    \label{fig:3}
\end{figure*}

\subsection{The role of the relativistic dispersion relation in DESP}
In order to optimize for the mass of the signal, we can formulate a physics-inspired optimization method that exploits the fact that eigenstates of the Topological Dirac equation satisfy the relativistic dispersion relation given by Eq.~\ref{eq:relativistic_dispersion_relation}. Therefore the reconstructed signal that more closely approximates an eigenstate of the Topological Dirac equation should minimize the relativistic dispersion relation error (RDRE) $\mathcal{S}_m$ over $m$. The RDRE  $\mathcal{S}_m$  is  given by \begin{equation}
    \mathcal{S}_m = |E_m^2-(\lambda_m^2+m^2)|
\end{equation}
where for any choice of $m$, $\lambda_m^2$ is the expectation of the reconstructed signal $\bm{\hat{\psi}}$ over the Laplacian      and $E_m$ is the expectation of the same signal over  the Hamiltonian, given by:
    \bea
\lambda_m^2=\frac{\hat{\bm\psi}_m^{\top}{\bf D}^2\hat{\bm\psi}_m}{\|\hat{\bm\psi}_m\|^2},\quad 
E_m=\frac{\hat{\bm\psi}_m^{\top}\bm{\mathcal{H}}\hat{\bm\psi}_m}{\|\hat{\bm\psi}_m\|^2}.
 \eea
Thus, optimizing $m$ according to the RDRE entails finding the value of the mass $m$ that minimizes:
\begin{equation}
\begin{split}
    \mathcal{S}_m & =|E_m^2-(|\lambda_m|^2+m^2)|\\&=\left|\frac{\left(\hat{\bm\psi}_m^{\top}({\bf D}+m\bm\gamma)\hat{\bm\psi}_m\right)^2}{\|\hat{\bm\psi}_m\|^4}-\frac{\hat{\bm\psi}_m^{\top}{\bf D}^2\hat{\bm\psi}_m}{\|\hat{\bm\psi}_m\|^2}-m^2\right|, 
\end{split}
 \end{equation}
where  $\mathcal{S}_m\geq 0$ in general and equal to zero if and only if $\hat{\bm\psi}_m$ is an eigenvector of the Dirac equation.

We observe that optimizing the loss function $\mathcal{L}_m$ given by Eq.~\ref{eq:loss_m}  in general gives different results with respect to the ones obtained by minimizing the RDRE $\mathcal{S}_m$. However, as long as the noise is not too high, the difference in the error made in reconstructing the true signal remains low (see Figure $\ref{fig:3}$).

\subsection{The improved performance of DESP}

\begin{figure*}[hbt!]
    \centering
    \includegraphics[width=0.99\textwidth]{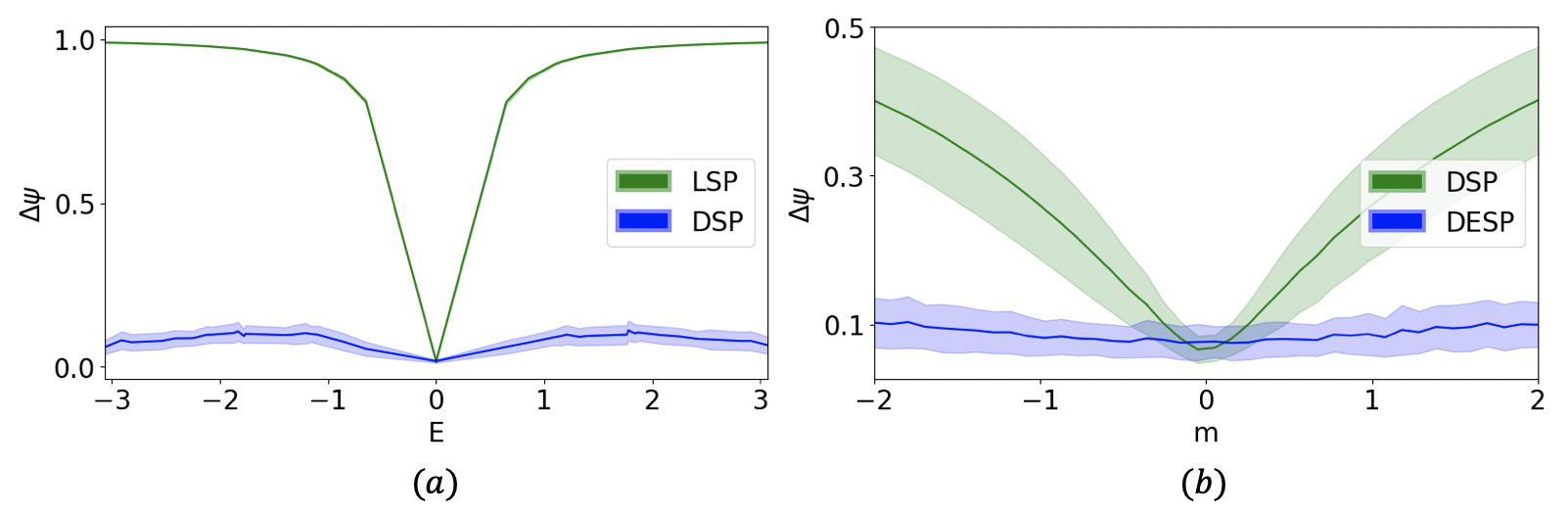}
    \caption{
    The DESP includes both the LSP and the DSP as subcases, and in general, can outperform both LSP and DSP. In order to compare the methods, we consider first a true signal given by an eigenstate of the Topological Dirac equation with $m=0$ and tunable value of the energy $E$ (indicating in this case the eigenvalue of the Dirac operator where $E=0$).
    By assuming that the value of the mass $m=0$ is known, DESP reduces to DSP that outperforms LSP (panel (a)) if the signal deviates from an almost harmonic signal (larger values of $|E|$). Indeed the error $\Delta \bm\psi$ of the reconstructed signal is much lower for the DSP than for the LSP for larger values of the energy $E$. 
    Secondly, we consider a true signal given by an eigenstate of the Topological Dirac equation with a tunable value of the mass $m$ and random value of the energy $E$. We show that DESP outperforms DSP by learning the true value of the mass, and the improvement in the error level $\Delta \bm\psi$ is more significant as the absolute value of the mass $m$ becomes larger (panel (b)).
    Here the results are obtained by considering $100$ noisy signals (the amplitude of the shaded regions indicates standard deviations) on the NGF network shown in Figure 1 with noise level $\alpha = 0.3$. The DESP uses the loss function $\mathcal{L}$ with parameter $\tau =10$ to infer the true mass. }
    \label{fig2}
\end{figure*}

\begin{figure*}[hbt!]
    \centering
    \includegraphics[width=0.99\textwidth]{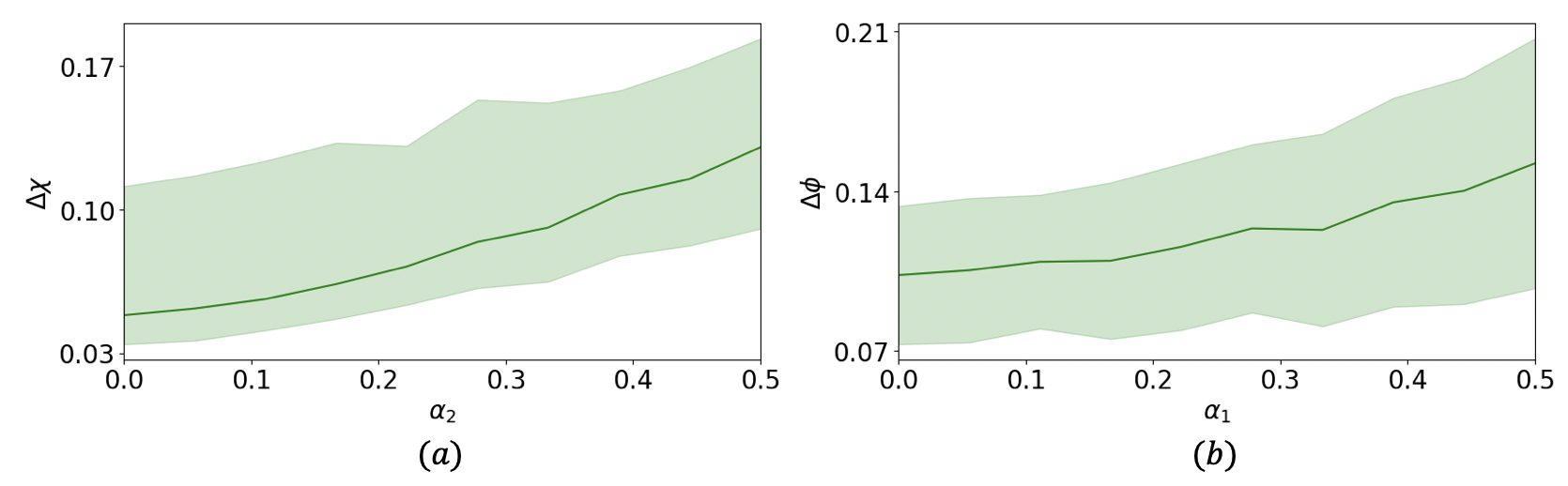}
    \caption{DESP can jointly process signals on nodes and edges and allow us to extract relevant information across topological signals of different dimensions. Here we plot the error $\Delta \bm\chi=\|\hat{\bm\chi}-\bm\chi\|/\|\bm\chi\|$, where $\hat{\chi}$ is the reconstructed node signal, and $\bm\chi$ is the true node signal as a function of the noise level $\alpha_2$ associated to the edge signal. We show that the error $\Delta \bm\chi=\|\hat{\bm\chi} \|$ made on the reconstruction of the node signal decreases as $\alpha_2$ is lowered, when the noise level on the edge signal is kept equal to $\alpha_2=0.5$ (panel(a)). Similarly, we show the error $\Delta \bm\phi=\|\hat{\bm\phi}-\bm\phi\|/\|\bm\phi\|$ where $\hat{\phi}$ is the reconstructed edge signal, and $\bm\phi$ is the true edge signal as a function of the noise level $\alpha_1$ on the node signal. Also in this case we show that the error  $\Delta \bm\phi$ made on the reconstruction of the edge signal decreases as the noise level $\alpha_1$ associated with the node signal is lowered, when the noise level on the edge signal is kept equal to $\alpha_2=0.5$ (panel (b)). 
    The results suggest an improvement in performance when the noise level on either nodes or links is independently reduced.
     The shaded area refers to the standard deviation error of  DESP calculated over $200$ noisy signals of the NGF network shown in Figure 1. In both panels, true signal is an eigenstate of the Topological Dirac equation with true mass $m=1.5$ and true energy $E = -3.31.$}
    \label{fig:enter-label}
\end{figure*}

The DESP algorithm reduces for $m=E=0$ to LSP and for $m=0$ to DSP. Therefore, the DESP algorithm can only provide an improved performance with respect to the two previous algorithms. In order to compare DESP with DSP and LSP and assess the entity of the improved performance of DESP, we consider the error in the reconstructed signal generated by the three algorithms when the true signal is aligned to a single eigenstate of the Topological Dirac Equation (see Figure \ref{fig2}).
We show that when the eigenstate is associated to energy $E$ and mass $m=0$, DSP can outperform LSP, in particular when the energy $E$ deviates significantly from zero. Thus also DESP can greatly outperform LSP in this case. When the eigenstate is an arbitrary eigenstate associated to energy $m$ and an arbitrary value of the energy $E$, DESP can also outperform DSP.
This is a great indication that DESP constitutes an important step forward in processing general node and edge topological signals.  
Note that, while here we work under the assumption that the true signal is aligned to a single eigenvector of the Topological Dirac equation, in the next section we will also address this limitation by formulating the Iterated Dirac-equation signal processing (IDESP) algorithm.
When validating the performance of the DESP algorithm, it is also important to answer the question whether jointly filtering node and edge signals can be beneficial to extract more information from data.
In order to address this question, we have considered the scenario where the noise level over node and edge signal is different and parametrized respectively by the parameters $\alpha_1$ and $\alpha_2$ (see Methods for details).
In particular, we have considered the error made by DESP on the reconstruction of the node signal $\Delta\bm\chi$ when the noise on the edge signal is decreased, showing that a less noisy edge signal can contribute to reconstruct better the edge signal (see panel (a) of Figure $\ref{fig:enter-label}$).
Similarly we have shown that the error made by DESP on the reconstruction of the edge signal $\Delta\bm\phi$ when the noise on the node signal is decreased, showing that a less noisy node signal can contribute to reconstruct better the node signal (see panel (b) of Figure $\ref{fig:enter-label}$).
These results indicate clearly that jointly processing node and edge signals can allow to extract more information from data, leveraging on the information content encoded by both node and edge signals.

\section{Iterated Dirac-equation signal processing (IDESP)}
\label{sec:IDESP}
\begin{figure*}[hbt!]
    \centering
    \includegraphics[width =0.99\textwidth]{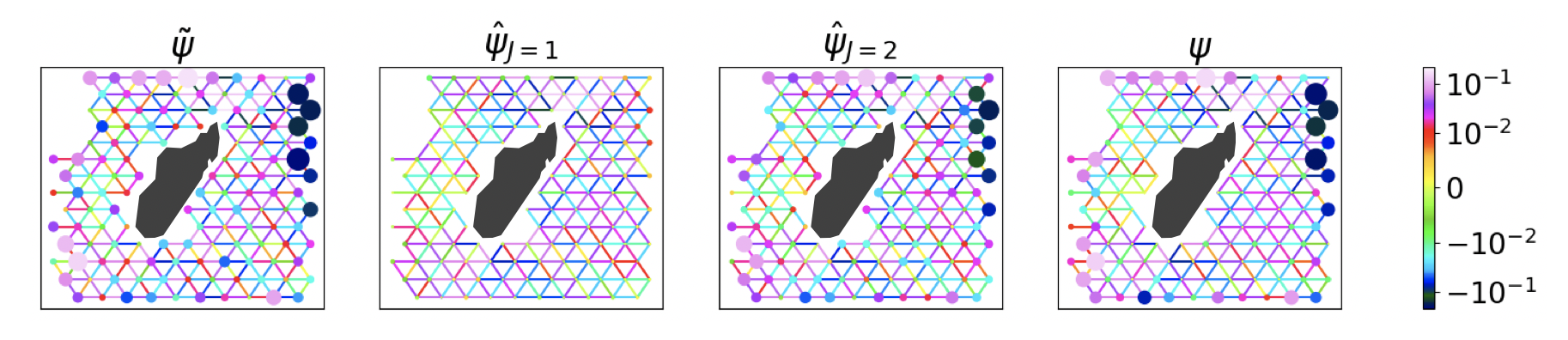}
    \caption{Visualization of IDESP$_\mathcal{S}$ applied to the drifter data around Madagascar. The figure illustrates, from left to right, the noisy signal (with noise level $\alpha = 0.25,$) the result of the first iteration of the IDESP ($J=1$), the final output of IDESP ($J=2$), and the real true signal. The real data has been analyzed and preprocessed previously in Refs.~\cite{schaub2020random,calmon2023dirac} and is freely available at \cite{gin_repository}}
    \label{fig_buoys_viz}
\end{figure*}

For treating real data, we need to go beyond our hypothesis that the true signal is a single eigenstate of the topological Dirac equation. Indeed in general, the true signal in real data will be a linear combination of different eigenstates of the Topological Dirac equation. Therefore, Algorithm \ref{alg:DESP} can only provide a prediction of the primary eigenstate $\boldsymbol{\hat{\psi}}_1$. However, we can iterate the algorithm on $\boldsymbol{\tilde{\psi}} - \boldsymbol{\hat{\psi}}_1$ to get the secondary eigenstate and we can iterate the process until the reduction of the coefficient of variation to the true or the estimated true value. This leads to the Iterated Dirac-equations signal processing (IDESP) algorithm \ref{alg:IDESP}, in which the DESP algorithm is iterated  $J$ times, providing the reconstructed signal 
\begin{equation}
    \boldsymbol{\hat{\psi}} = \sum_{j=1}^J \boldsymbol{\hat{\psi}}_j.
    \label{rec_iterated}
\end{equation}
However, iterating the DESP algorithm is not enough as we need reliable criteria for determining when to stop iterating it. Indeed,  increasing the number of iterations  $J$ may not always lead to an an increase in accuracy, as after a certain number of iterations, we might end up reconstructing also part of the noise. 
In the following, we assume that the true coefficient of variation (noise-to-signal ratio)  $c_V^{\textrm{true}}$ of the measured signal, given by 
\bea
c_V^{\textrm{true}}=\frac{\|\bm\psi-\boldsymbol{\tilde{\psi}}\|_2}{\|\bm\psi\|_2}
\eea is either known or reliably estimated. 
In this case, the Iterated Dirac-equation signal processing (IDESP) algorithm will iterate the DESP process up to the iteration $J_{\textrm{opt}}$ that minimizes the absolute difference of the coefficient of variation of the reconstructed signal and the true coefficient of variation.
Specifically, the IDESP will stop for $J=J_{\textrm{opt}}$ with 
\begin{equation}
     J_{\textrm{opt}} =\mbox{argmin}_J |c_V(J)-c_V^{\textrm{true}}|.
     \label{eq:cv_opt}
\end{equation}
where the {\em coefficient of variation} $c_V(J)$ of the reconstructed signal after the $J$ iterations is given by 
\begin{equation}
  {c_V}(J)=\frac{\|\sum_{j=1}^J \boldsymbol{\hat{\psi}}_j - \boldsymbol{\tilde{\psi}}\|_2}{\|\sum_{j=1}^J \boldsymbol{\hat{\psi}}_j\|_2}, 
\end{equation}
Only in this way, we have that if the reconstructed signal is equal to the true signal, ${c_V}(J)$ is the true noise-to-signal ratio and thus we guarantee that our optimization criterion given by Eq.~\ref{eq:cv_opt} effectively stops at the right place.
In the scenario in which the true coefficient of variation is not known, this algorithm can always be used to provide an ensemble of signal reconstructions, i.e.~providing for any possible value of $c_V^{\textrm{true}}$ the reconstructed signal $ \boldsymbol{\hat{\psi}}$ given by Eq.~\ref{rec_iterated} with  $J=J_{\textrm{opt}}(c_V^{\textrm{true}})$.

\begin{algorithm}[H]
	\caption{Iterated Dirac-Equation Signal Processing (IDESP) }
	\label{alg:IDESP}
	\begin{algorithmic}[1] 
 \Statex Input:  All the required inputs for DESP Algorithm including the measured signal $\tilde{\bm\psi}$; the estimated or true coefficient of variation $c_V^{\textrm{true}}$ of the measured signal $\tilde{\bm\psi}$.
 \Statex Output: The output of the IDESP Algorithm starting from the measured signal $\tilde{\bm\psi}$ is indicated as $\hat{\bm\Psi}=\mbox{IDESP}(\tilde{\bm\psi})$
 \State $\Delta_0 c\leftarrow 0$
  \State $J \leftarrow 1$
  \State $\bm\phi \leftarrow \tilde{\bm\psi}$
 \State $\boldsymbol{\hat{\psi}}_1 \leftarrow  \mbox{DESP}({\bm\phi})$  
 \State $c_V(1)\leftarrow{\|{\boldsymbol{\hat{\psi}}}_1 - \boldsymbol{\tilde{\psi}}\|_2 }/{\| {\bm{\hat{\psi}}}_1\|_2}$
\State  $\Delta_1 c\leftarrow \left| c_V(1)- c_V^{\textrm{true}} \right|$
\While{$\Delta_J c> \Delta_{J-1} c$}
 \State $J \leftarrow J + 1$
 \State $\bm\phi \leftarrow \tilde{\bm\psi}-\sum_{j=1}^{J-1}\hat{\bm\psi}_j$
 \State $\boldsymbol{\hat{\psi}}_{J} \leftarrow \mbox{DESP}(\bm\phi)$  
\State $c_V(J)\leftarrow{\|\sum_{j=1}^J{\boldsymbol{\hat{\psi}}}_j - \boldsymbol{\tilde{\psi}}\|_2 }/{\| \sum_{j=1}^J{\bm{\hat{\psi}}}_j\|_2}$
\State  $\Delta_J c\leftarrow \left| c_V(J)- c_V^{True} \right|$
            \EndWhile
            \State  $\boldsymbol{\hat{{\Psi}}}=\sum_{j=1}^J \boldsymbol{\hat{\psi}}_j$.
	\end{algorithmic}
\end{algorithm}

\begin{figure}[!htb]
\centering
    \includegraphics[width=0.95\columnwidth]{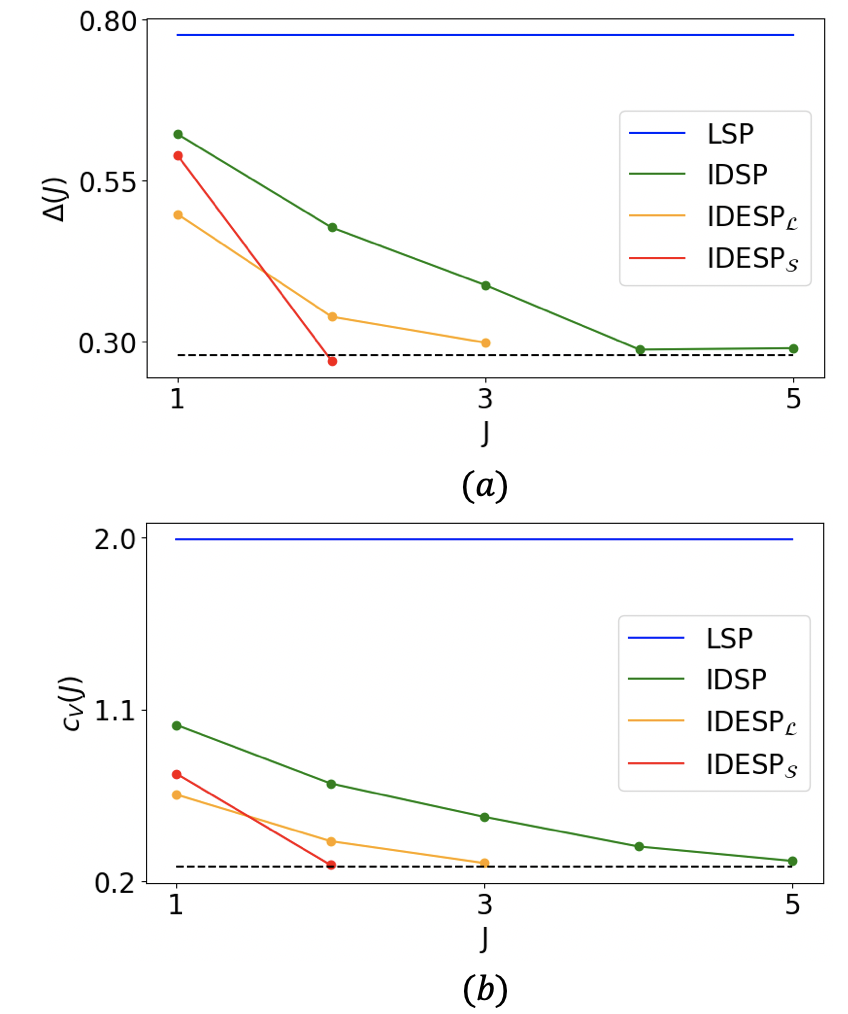}
    \caption{The performance of the IDESP algorithm implementing the optimization of the mass parameter with the loss function (IDESP$_{\mathcal{L}}$) or with the RDRE (IDESP$_{\mathcal{S}}$) on real drifters data around Madagascar is compared to the IDSP and to the LSP at each iteration $J$. Panel (a) displays the true error $\Delta(j)$ on the reconstructed signal at iteration $J$, panel (b) displays the coefficient of variation $c_V(J)$.
    The result of the LSP does not depend on the iteration, the results of DSP is given by the result of IDSP for $J=1$.
    The true coefficient of variation of the data is shown as a dotted line.
    The noise level is $\alpha=0.25,$ and the parameter $\tau$ is given by $\tau=15$.}
    \label{fig_buoys}
\end{figure}

We test the IDESP on the real dataset of drifters in the ocean from the Global Ocean Drifter Program available at the AOML/NOAA Drifter Data Assembly Center  already analyzed in Ref.~\cite{schaub2020random,calmon2023dirac}    (data available at the Repository\cite{gin_repository} see Methods for details), finding fairly good results (see  Figure \ref{fig_buoys_viz} for a visualization of the performance of the IDESP algorithm).
In order to  quantify the performance of the IDESP on this real dataset,  in  Figure \ref{fig_buoys} we monitor the true error $\Delta(J)$ at iteration $J$ of the algorithm, i.e.
\bea
\Delta(J)=\left\|\sum_{j=1}^J \boldsymbol{\hat{\psi}}_j - \boldsymbol{{\psi}}\right\|_2.
\eea
We observe that the error lowers up to $J=J_{\textrm{opt}}$, validating the performance of the adopted IDESP algorithm.
Due to the nature of the signal, IDESP can offer a great improvement. Note that this improvement can be observed not only when in the DESP algorithm we determine  the mass by minimizing the loss $\mathcal{L}$ but also when we determine the mass by minimizing the RDRE $\mathcal{S}$. 
The iterated procedure can be also be applied to the DSP algorithm leading to the Iterated Dirac signal processing (IDSP) algorithm finding very significant improvements as well, however using the IDESP allows to achieve the same coefficient of variations with fewer iterations, indicating the better suitability of the IDESP in approximating the true signals.

\section{Conclusions}
In this work, we propose Dirac-equation signal processing (DESP), a physics inspired algorithm that leverages on the properties of the Topological Dirac equation to filter and process jointly node and edge signals defined on the same network.
We have demonstrated through both theoretical insights and numerical verification on synthetic and real data that DESP reduces to the previously proposed LSP and DSP and that in general scenarios can outperform both of them. In particular, DESP allows to jointly process both node and edge signals, extracting relevant information across the topological signal of different dimensions, adaptively adjusting for their different scales thanks to the introduction of the learnable mass parameter $m$.
While the DESP processes signals assuming they are formed by a single eigenstate of the Topological Dirac equation, the IDESP allows to treat more general signals formed by a linear combination of eigenstates of the Topological Dirac equation. This latter algorithm can further boost the performance of DESP on real signals as demonstrated here by applying this algorithm to an extensive dataset of drifters around Madagascar.

We hope that these results will raise further interest into the use of the Topological Dirac operator and the Topological Dirac equation in Artificial intelligence, stimulating further research in both signal processing and neural networks. For instance, in signal processing, an open question is to filter topological signals across a multiplex network or knowledge graph formed by networks of networks, thus exploiting the relevant information in the different layers without simply aggregating the data. Although the focus of this paper is on topological machine learning, it is noteworthy that the Dirac operator by jointly processing node and edge signals could improve the long-range information from distant nodes and therefore ameliorate over-squashing and over-smoothing problems of topological deep learning found in graph neural networks.

\section*{Methods}
\subsection*{Noise model}
In DESP the noise $\bm\epsilon$ associated to the noise level $\alpha$ is generated as follows.
First we draw the vector ${\bf x}$ of i.i.d.~Gaussian variables $x_{\hat{\sigma}}$ with average zero and standard deviation $\alpha$, associated to each simplex $\hat{\sigma}$ of the network (node or edge) i.e.~$x_{\hat{\sigma}} \sim \mathcal{N}(0,\alpha)$ and then we filter out their harmonic component, putting 
\bea
\bm\epsilon=\frac{{\bf D}{\bf D}^{+}{\bf x}}{\sqrt{D}},
\label{eq:epsilon}
\eea
where ${\bf D}^{+}$ indicates the pseudo-inverse of the Dirac operator and  $D$ its rank.
This is the same noise model adopted in for DSP in Ref.~\cite{calmon2023dirac}.
In Figure \ref{fig:enter-label}, we consider a variation of this noise model in which the vector ${\bf x}$ is formed by i.i.d. Gaussian variables $x_{\hat{\sigma}}$ with different standard deviations depending on the dimension of the simplex $\hat{\sigma}$. In particular we associate the nodes with a noise of standard deviation $\alpha_1$, i.e. $x_{r}\sim \mathcal{N}(0,\alpha_1)$  and the edges with standard deviation $\alpha_2$, i.e. $x_{[rs]}\sim \mathcal{N}(0,\alpha_2)$. The noise $\bm\epsilon$ is then given by Eq.~\ref{eq:epsilon}.

\subsection*{Drifter dataset}
We test the  IDESP algorithm on the real dataset of drifters in the ocean from the
Global Ocean Drifter Program available at the AOML/NOAA Drifter Data Assembly Center~\cite{drifter}. The drifters data set already analyzed in Ref.~\cite{schaub2020random,calmon2023dirac} consists of the individual trajectories of $339$ buoys around the island of Madagascar in the Pacific Ocean. Projected onto a tessellation of the space, this yields $339$ edge-flows, each representing the motion of a buoy between pairs of cells  (data available at the Repository\cite{gin_repository}. The resulting network is formed by  $N_0=133$ nodes, and $N_1=322$ links. The edge topological signal $\bm \theta$ is given on each edge by the sum of all the  $339$ trajectories passing through that edge, representing the net physical flow along each edge. In the absence of a true node signal, we generate a non-trivial topological spinor playing the role of our true signal $\bm\psi$ from the exclusive knowledge of the edge signal $\bm\theta$. 
Specifically, we consider the topological signal $\bm \sigma=({\bf 0},\bm\theta)$  defined on both nodes and edges and we put 
\bea
\bm\psi =C(\bm \sigma+{\bf D} \bm \sigma),
\label{eq:dr}
\eea
where $C$ is the normalization constant that enforces  $\|\bm\psi\|_2=1$.

\section*{Acknowledgments}
The authors would like to thank the Isaac Newton Institute for Mathematical Sciences, Cambridge, for support and hospitality during the programme Hypergraphs: Theory and Applications, where work on this paper was undertaken. This work was supported by EPSRC grant EP/R014604/1  and  partially supported by  grants from the Simons Foundation (Y.T.~and G.B.)
\bibliography{refs}

\end{document}